\documentclass[aps,prl,superscriptaddress,unsortedaddress,twocolumn,preprintnumbers,amsmath,amssymb]{revtex4-1}
\usepackage{graphicx}
\usepackage{dcolumn}
\usepackage{bm}
\usepackage{upgreek}
\usepackage{txfonts}
\usepackage{color}
\newsavebox{\tempbox}

\begin{document}

\title{Mechanism of generation of a Josephson phase shift by an Abrikosov vortex}

\author{T. Golod$^1$}
\author{A. Pagliero$^1$}
\author{V. M. Krasnov$^{1,2}$}
\email{Vladimir.Krasnov@fysik.su.se}

\affiliation{$^1$ Department of Physics, Stockholm University,
AlbaNova University Center, SE-10691 Stockholm, Sweden }
\affiliation{$^2$ Moscow Institute of Physics and Technology,
State University, 9 Institutsiy per., Dolgoprudny, Moscow Region
141700 Russia}

\date{\today}

\begin{abstract}

Abrikosov vortex contains magnetic field and circulating currents
that decay at a short range $\lambda \sim 100$ nm. However, the
vortex can induce a long range Josephson phase shift at distances
$r\sim\mu$m$\gg\lambda$. The mechanism of this puzzling phenomenon
is not clearly understood. Here we present a systematic study of
vortex-induced phase shift in planar Josephson junctions. We make
two key observations: (i) The cutoff effect: although
vortex-induce phase shift is a long-range phenomenon, it is
terminated by the junction and does not persists beyond it. (ii) A
crossover from linear to superlinear dependence of the phase shift
on the vortex polar angle occurs upon approaching of the vortex to
the junction. The crossover occurs at a distance comparable with
the penetration depth. This, together with theoretical and
numerical analysis of the problem, allows unambiguous
identification of two distinct and independent mechanisms. The
short range mechanism is due to circulating vortex currents {\it
inside} superconducting electrodes without involvement of magnetic
field. The long range mechanism is due to stray magnetic fields
{\it outside} electrodes without circulating vortex currents. We
argue that understanding of controlling parameters of
vortex-induced Josephson phase shift can be used for development
of compact and fast electronic devices with low dissipation power.

\end{abstract}


\maketitle

\section{I. Introduction}

Josephson electronics operates with the quantum-mechanical phase
difference $\varphi$ between two superconducting electrodes. In
conventional Josephson junctions the phase difference is zero in
the absence of applied current and is spatially independent in the
absence of external magnetic field. It is anticipated that
unconventional junctions either with a fixed phase shift
$\varphi\neq 0$ ($\varphi$-junctions), or with an in-built spatial
phase variation along the junction $\Delta\varphi$ ($0-\varphi$
junctions) may provide additional functionality in Josephson
electronics
\cite{Blatter_2001,Ortlepp_2006,Feofanov_2010,Soloviev_2017}. For
example, they can operate as autonomous and persistent phase
batteries - one of key elements of quantum
\cite{Blatter_2001,Martinis_2008,Gershenson_2009,Mooij_2002,Wulf_2006,Lukens_2007,Mooij_2009,Semba_2010,Mooij_2011,Schwarz_2013,Deng_2015}
and digital Josephson electronics
\cite{Ortlepp_2006,Balashov_2007,Dimov_2008,Feofanov_2010}. Such
junctions can be also used for development of novel types of
cryogenic memory
\cite{Goldobin_2013,Zdravkov_2013,Dresselhaus_2014,Golod_2015,Bakurskiy_2016,Birge_2018}.
The $\pi$-phase shift is most commonly needed, e.g., for bringing
the flux-qubit to the degeneracy point
\cite{Blatter_2001,Martinis_2008,Gershenson_2009,Mooij_2002}, for
realization of complementary digital Josephson electronics
\cite{Ortlepp_2006,Balashov_2007,Dimov_2008,Feofanov_2010} and for
maximum distinction between 0 and 1 state in a memory cell
\cite{Golod_2015}.

Several ways of making $\varphi$ and $0-\varphi$ junctions are
known. The spatial phase variation within the junction can be
introduced by inhomogeneity, e.g. uneven distribution of critical
or bias current, which will cause the self-field effect
\cite{Geshkenbein_1992,Krasnov_1997,Gaber_2005,Linder_2013,Golod_2019}
and affect dynamics of Josephson vortices
\cite{Geshkenbein_1992,Krasnov_1997,Chesca_2017}. Junctions with a
fixed $\varphi=\pi$ phase can be realized using hybrid
Superconductor/Ferromagnet (S/F) structures
\cite{Ryazanov_2001,Aprili_2002,Calemczuk_2004} and the
unconventional sign-reversal order parameter $d$-wave
\cite{Ortlepp_2006,VanHarlingen_1995,Tsuei_2000} or $s_{\pm}$
\cite{Ota_2009,Linder_2009,Koshelev_2012,Kalenuyk_2018}. Also
$\varphi$-junctions with arbitrary phase can be realized using
inhomogeneous S/F junction \cite{Koshelev_2003,Goldobin_2012}, or
junctions with a strong spin-orbit coupling \cite{Bergeret_2015}.
The above mentioned phase-shifted junctions are either persistent
but not tunable (e.g. SFS junctions and junctions with
sign-reversal order parameter), or tunable but not persistent
(e.g. $0-\varphi$ junctions based on uneven current injection
\cite{Gaber_2005}).

An Abrikosov vortex (AV), having a $2\pi$ phase rotation, can
induce a phase shift in a nearby Josephson junction (JJ)
\cite{Golod_2010,Gurevich_1994,Aslamazov_1984,Clem_2011,Kogan2014,Mironov_2017}.
The vortex-induced phase shift depends on the distance and
geometry \cite{Golod_2010,Clem_2011}. Since vortices can be easily
and controllably manipulated (displaced, introduced or removed) by
magnetic field \cite{Golod_2010,Milosevic_2009}, current
\cite{Golod_2015,Finnemore_1994,Milosevic_2009}, light or heat
\cite{Tamarat_2016,Mironov_2017} pulses, they can be used for
creation of memory cells\cite{Golod_2015} and tunable but
persistent phase batteries. However, the mechanism of
vortex-induced Josephson phase shift is still not well understood.
Although circulating currents and magnetic fields of the AV decay
in a superconductor exponentially at a short range determined by
the London penetration depth  $\lambda \sim 100$ nm, the
vortex-induced phase shift is decaying much slower and persists at
distances of few microns \cite{Golod_2010}. To some extent it
resembles the long-range Aharonov-Bohm effect \cite{AB_1959}. For
example, there is a seeming direct relation between the
vortex-induced phase shift and rotation of the phase in the vortex
(polar angle $\Theta_v$) within the junction \cite{Golod_2010}.
Yet, it can not be the Aharonov-Bohm effect because the phase
difference at an interval (junction length) is not a
gauge-invariant quantity and has no physical significance. At
present there is no quantitative understanding of vortex-induced
Josephson phase shift for realistic sample geometries.

Here we perform a systematic analysis of vortex-induced Josephson
phase shifts in Nb-based planar Josephson junction with artificial
vortex traps. We make two key observations: (i) The cutoff effect:
although vortex-induce phase shift is a long-range phenomenon, it
is terminated by the junction and does not persists beyond it.
(ii) A crossover from linear to superlinear dependence of the
phase shift on the vortex polar angle occurs upon approaching of
the vortex to the junction. The crossover occurs at a distance
comparable with the penetration depth. This allows unambiguous
identification of two distinct and independent mechanisms. The
short range mechanism is due to circulating vortex currents {\it
inside} superconducting electrodes without involvement of magnetic
field. The long range mechanism is due to stray magnetic fields
{\it outside} electrodes without circulating vortex currents. It
is long-range because stray fields can not enter the
superconductor and have to spread-out along the surface until the
electrode edge up to arbitrary long distances. Our conclusions are
supported by theoretical analysis and numerical simulations. We
argue that the detailed understanding of geometrical factors
controlling the vortex-induced Josephson phase shift facilitates
development of compact and fast electronic devices with low
dissipation power.

\begin{figure*}[t]
    \includegraphics[width=0.8\textwidth]{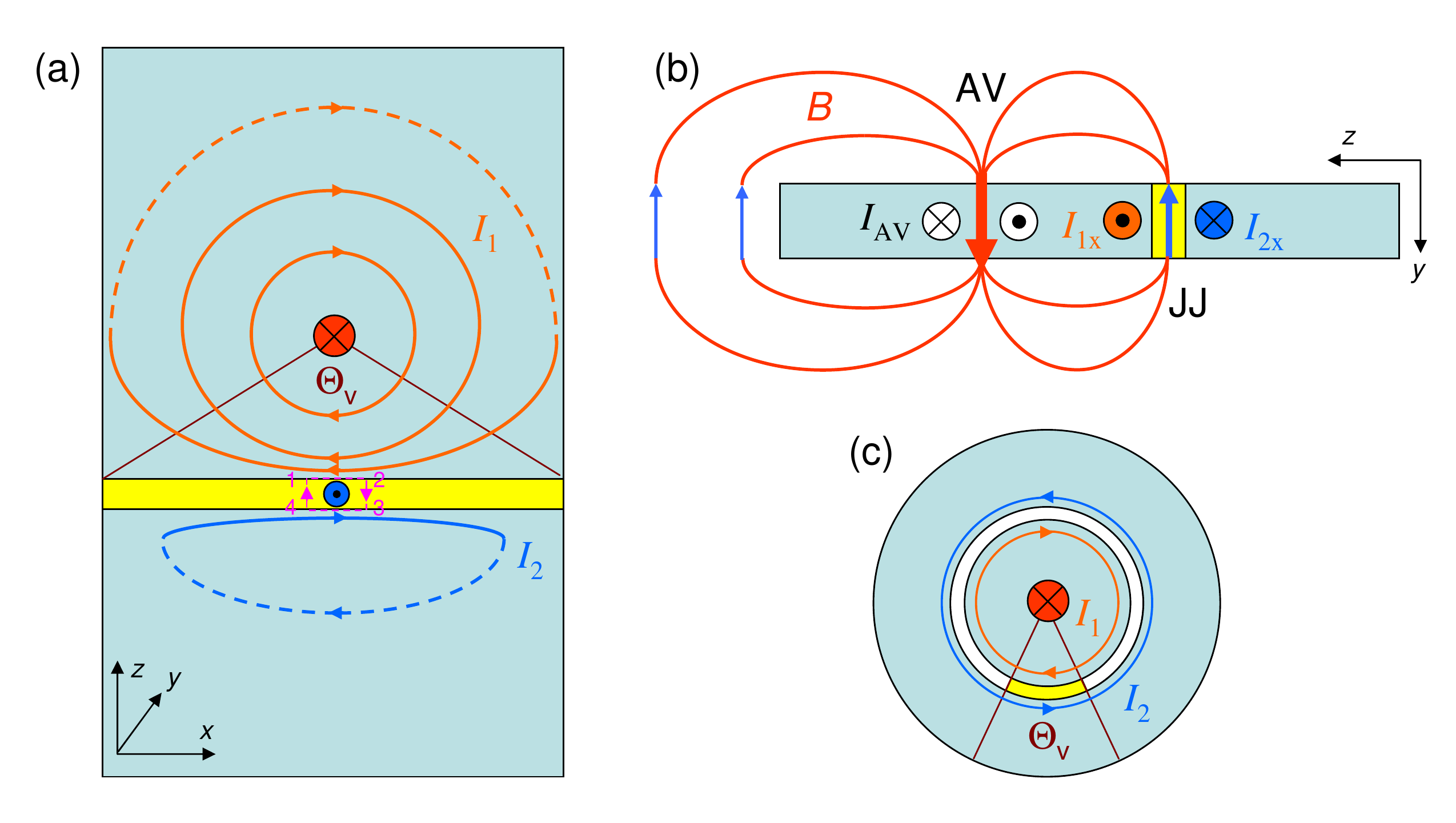}
     \caption{\label{fig:T1} (Color online). Sketches of considered vortex-junction configurations.
     (a) A top view of a planar junction (yellow region) with an Abrikosov vortex in the first (top) electrode.
     Lines indicate current streamlines.
     (b) A side-view of the same configuration. Lines indicate magnetic field lines of vortex stray fields.
     (c) A top view of Corbino disk geometry, in which the fraction of stray flux entering the junction is proportional to the polar
     angle of the vortex within the junction $\Theta_v$.}
\end{figure*}

\section{II. Theoretical analysis}

Let's consider a single vortex in one electrode of a planar
junction, as sketched in Fig. \ref{fig:T1} (a). We assume that the
junction is short, i.e., its length $L_x$ is smaller than double
the Josephson penetration depth $\lambda_J$. The vortex has a
$2\pi$ phase rotation, clockwise circulating currents and magnetic
flux, decaying at some characteristic scale $\lambda^*$. In the
limiting cases of very thick films, $d\gg\lambda$, the decay
length {\em inside} a superconducting electrode is equal to the
London penetration depth, $\lambda^*=\lambda$, and for very thin
films, $d\ll\lambda$ to the Pearl length $\lambda^*=\lambda^2/d$
\cite{Pearl_1964}. Finally, there are stray fields from the vortex
{\em outside} electrodes. They can not enter superconducting
electrodes and, therefore, stretch along the surface until the
edges, where they eventually close, as sketched in Fig.
\ref{fig:T1} (b). In this section we discuss how do four
vortex-related factors, i) phase rotation of the superconducting
condensate, ii) circulating current, iii) inner field and iv)
stray fields, contribute to the Josephson phase shift.

As reported earlier \cite{Golod_2010}, the value and the sign of
AV-induced Josephson phase shift $\Delta \varphi_v$ can be
obtained using a naive assumption of a rigid $2\pi$ phase rotation
around the vortex (the London gauge). In this case the phase shift
is just equal to the polar angle,
\begin{equation}\label{AQ1}
\varphi_{v}(x)= -V \arctan \left(\frac{x-x_v}{|z_v|}\right),
\end{equation}
where $V$ is the vorticity (+1 for a vortex, -1 for an
antivortex). Here $x_v$ is the coordinate of the vortex along the
junction and $z_v$ is the distance to the junction.
The total phase shift is equal to the polar angle of the vortex
within the junction $\Theta_v$, indicated in Fig. \ref{fig:T1}
(a), $\Delta \varphi_v=\Theta_v =\arctan
[x_v/|z_v|]+\arctan[(L_x-x_v)/|z_v|]$. When vortex is close to the
junction, $|z_v|\ll L_x$, it induces a step-like $\pi$ shift,
turning the junction into a $0-\pi$ state
\cite{Golod_2010,Golod_2015}. The sign of the phase shift is
determined by the direction of phase rotation. In Fig.
\ref{fig:T1} (a) we sketch the case of a vortex (clockwise
rotation) in the top electrode. Since the Josephson phase is the
difference in phases of top and bottom electrodes, a vortex
induces a negative phase shift $\Delta \varphi<0$, irrespective in
which electrode it is placed. Indeed, when it is placed in the top
electrode, it induces a negative phase gradient solely in the top
electrode. But if we translate it to the bottom electrode it will
induce a positive phase gradient solely in the bottom electrode.
This explains the minus sign and the absolute value $|z_v|$ in Eq.
(\ref{AQ1}). However, as we already mentioned, rigid phase
rotation can not be a proper explanation of the vortex-induces
Josephson phase shift because in quantum mechanics the phase
difference is determined only at a closed loop and can not acquire
a physically significant value at an interval (junction length).
Therefore, the phase shift has to be induced by circulating
currents and fields of the vortex.

Current densities $J_{1,2}$ in the two electrodes are described by
the generalized second London equation:
\begin{equation}\label{London}
J_{1,2}=\frac{c}{4\pi\lambda_{1,2}^2}
\left[\frac{\Phi_0}{2\pi}\triangledown\Theta_{1,2}-A\right].
\end{equation}
Here $\Theta$ and $A$ are the corresponding scalar (phase) and
vector potentials. The gauge-invariant Josephson phase difference
is obtained by integration of Eq.(\ref{London}) over an
infinitesimal contour covering the barrier, 1-2-3-4 in Fig.
\ref{fig:T1} (a),
\begin{equation}\label{Adphi}
\frac{1}{2\pi}\frac{\partial\varphi}{\partial x}=
\frac{4\pi\lambda_{1,2}^2}{\Phi_0
c}(J_x^{(1)}-J_x^{(2)})+\frac{t}{\Phi_0}B_y.
\end{equation}
Here $t$ is the width of the barrier, $J_x^{(1,2)}$ are
$x-$components of supercurrent densities in the vicinity of the
barrier in electrodes 1 and 2, $B_y$ is the $y$-component of
magnetic induction in the barrier.

In the Meissner state, $J_x^{(1,2)}$ are obtained by solving
Eq.(\ref{London}) in the electrodes, with boundary conditions $B =
H$ outside the junction, $z=\pm d_{1,2}$. Straightforward
calculations yield \cite{Fluxon_1997}:

\begin{equation}\label{Adphi0}
\frac{\partial\varphi_M}{\partial x}=\frac{2\pi}{\Phi_0} \left[B
\Lambda - H S \right].
\end{equation}

Here $\Lambda=t+\sum_{i=1,2}\lambda_i \coth\frac{d_i}{\lambda_i}$
and $S=\sum_{i=1,2} \lambda_i \textrm{cosech}
\frac{d_i}{\lambda_i}$. For a short junction screening within the
junction is negligible, $B=H$. This leads to proportionality of
the phase gradient to the applied field.
Integrating $\partial\varphi/\partial x$ over the junction length
$x\in [0,L_x]$ provides a simple relation between the total phase
shift within the junction $\Delta \varphi
=\varphi(L_x)-\varphi(0)$ and the flux:
\begin{equation}\label{AdphiF}
\Delta \varphi_M = 2\pi \frac{\Phi}{\Phi_0}.
\end{equation}
Here $\Phi=d_{eff}\int^{L_x}_0{B_y dx}$ is the total flux in the
junction and $d_{eff}$ is the effective magnetic thickness of the
junction. For conventional overlap junctions $d_{eff}=\Lambda-S
\simeq t+\lambda_1 + \lambda_2$ for thick electrodes
$d_{1,2}\gg\lambda$, and $\simeq t+d_1/2 + d_2/2$ for thin
electrodes $d_{1,2}\ll\lambda$, respectively. For planar
junctions, studied below, $d_{eff}$ is determined by the size
$L_x$ of junction electrodes.
\cite{Kogan2001,Mints2008,Clem2010,Boris_2013,Golod_2019}

In the presence of AV the Josephson phase difference is a
superposition of contributions from the Meissner state, $\varphi_M
(H)$, and vortex-induced phase shift at zero field,
$\varphi_V(0)$:
\begin{equation}\label{phi_tot}
\varphi=\varphi_M (H)+\varphi_V (0).
\end{equation}

Currents and fields contribute differently to the phase shift. For
example, in the mesoscopic case $L_x\ll\lambda^*$, the total flux
carried by the vortex $\Phi_v \sim
(L_x/\lambda^*)^2\Phi_0\ll\Phi_0$ becomes negligible
\cite{Fluxon_1997}. Thus, there is no magnetic field neither
inside, nor outside the vortex and current makes the only
contribution to the phase shift. All previous theoretical studies
considered a similar case
\cite{Aslamazov_1984,Gurevich_1994,Clem_2011,Kogan2014,Mironov_2017}.
We note that junctions studied here and earlier
\cite{Golod_2010,Golod_2015} have intermediate electrode thickness
$d \sim \lambda$ \cite{Boris_2013} and their sizes $L_x\sim
5-6~\mu$m are significantly larger that $\lambda^*\sim 100-300$
nm. Therefore, stray fields can not be neglected in the
experiment. Below we consider limiting cases of both small and
large junctions.

\begin{figure*}[t]
    \includegraphics[width=0.95\textwidth]{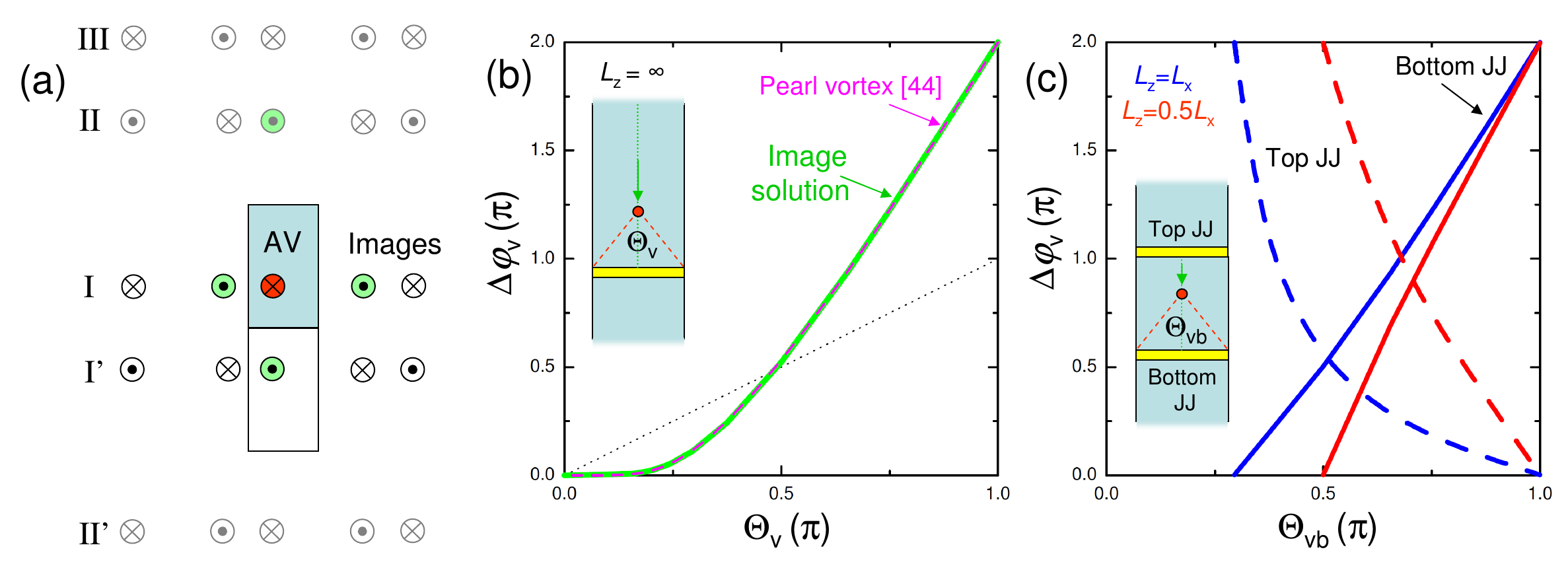}
     \caption{\label{fig:T2} (Color online). (a) Illustration of the image array solution for a rectangular electrode with a vortex.
     Mirror reflections from edges lead to an infinite series of vortices and antivortices. The total vortex-induced phase shift
     is the sum over the image array. (b) Calculated variation of the total phase shift versus vortex polar angle for a junction
     with narrow $L_x\ll \lambda^*$, long $L_z=\infty$ electrodes. Calculations are made for an
     antivortex approaching the junction along the middle line of the electrode,
     $x_v=L_x/2$, from $z_v=\infty$ ($\Theta_v=0$) to $z_v=0$ ($\Theta_v=\pi$).
     Green line represents the image array solution, which
coincides with the solution from Ref. \cite{Clem_2011} for a Pearl
antivortex (dashed magenta line). (c) Image-array solutions for
finite-size, rectangular-shape electrodes with $L_z=L_x$ (blue
lines) and $L_z=L_x/2$ (red lines). Here we consider a double
junction experiment and show variation of the total phase shift in
the bottom (solid lines) and top (dashed lines) junctions made at
the upper and lower edges of the electrode, versus the polar angle
in the bottom junction $\Theta_{vb}$. Calculations are made for an
antivortex moving along the middle line $x_v=L_x/2$ of the
electrode (vertical green doted line) from the top, $z_v=L_z$
(minimal value of $\Theta_{vb}$) to the bottom $z_v=0$
($\Theta_{vb}=\pi$). }
\end{figure*}

\subsection{II A. Current induced phase shift in a mesoscopic limit $L_x\ll\lambda^*$}

Let's first consider the case of a very small junction with sizes
$L_x\ll\lambda^*$. In this case the vortex does not carry flux
$\Phi\simeq 0$ and all magnetic field effects can be neglected.
Then the London equation \ref{London} can be rewritten in
cylindrical ($r,\Theta$) coordinates as
\begin{equation}\label{London_Pearl}
    rot J=\frac{c\Phi_0}{4\pi\lambda^{*2}}\delta(\Theta),
\end{equation}
where $\delta(\Theta)$ is a delta-function. Direct integration
yields
\begin{equation}\label{J_Pearl}
    J_{\Theta}=\frac{c\Phi_0}{8\pi^2\lambda^{*2}}\frac{1}{r}
\end{equation}
Since $B_y=0$, there are no field-induced screening currents in
the second electrode, $J_x^2 =0$. If we substitute Eq.
(\ref{J_Pearl}) into Eq. (\ref{Adphi}), we obtain
$\varphi_V=\Theta$, which is equivalent to Eq. (\ref{AQ1}).
However, this is not a self-consistent solution because Eq.
(\ref{J_Pearl}) does not take into account boundary conditions at
the edges of the electrode. The circulating current can not cross
the edges \cite{Note0}. This leads to distortion of stream lines,
as sketched in Fig. \ref{fig:T1} (a).

The boundary condition can be accounted for by using a well known
image method \cite{Gurevich_1994}. A mirror image with opposite vorticity should be
added at the opposing side from the edge so that it cancels out
the perpendicular component of currents through this edge.
However, it adds a smaller current at the opposite edge. To
compensate for it we should introduce an image of an image at the
opposing edge, and so on. This leads to an infinite series of
images due to reflections from all the edges, as illustrated in
Figure \ref{fig:T2} (a). Each image vortex generates a phase shift
$\varphi_v(x_n,z_m)$ according to Eq. (\ref{AQ1}) with image
vortex coordinates $(x_n,z_m)$ and corresponding vorticity
$V_{nm}$. The image solution for vortex-induced phase shift is
then
\begin{equation}\label{ImageSum}
\varphi_v=\sum_{n,m}{\varphi_v(x_n,z_m)}.
\end{equation}
Here the sum is taken over the image array. It is rapidly converging and can be easily
calculated, as described in the Appendix A. For the case of
semi-infinite electrode, $L_z=\infty$, the image array consists of
the primary image row-I due to reflections from left and right
edges, and an anti-row I' due to mirror reflection from the bottom
(junction) edge, see Fig. \ref{fig:T2} (a). The green line in Fig.
\ref{fig:T2} (b) shows the corresponding total antivortex induced
phase shift as a function of the polar angle of the vortex within
the junction $\Theta_v$, for a vortex in the middle of electrode
$x_v=0.5 L_x$.

In Ref.\cite{Clem_2011} J. Clem obtained a self-consistent
solution for a Pearl- vortex-induced phase shift in a thin film
planar Josephson junction with narrow long electrodes
($d\ll\lambda$, $L_x\ll\lambda_P$, $L_z=\infty$) using a conformal
mapping method technique. The solution is described in Appendix B.
It is shown by the dashed magenta line in Fig.
\ref{fig:T2} (b). Apparently it coincides with the image array
solution. The coincidence is not occasional because
in the considered flux-free mesoscopic case the only vortex
feature is the delta-function phase singularity,
Eq.(\ref{London_Pearl}), irrespective of the vortex type (Pearl or
Abrikosov). Consequently, the solutions are also identical.

From Fig. \ref{fig:T2} (b) it is seen that a vortex far away from
the junction, $\Theta_v\simeq 0$, induces a negligible phase shift
$\Delta\varphi\simeq 0$. Upon approaching the junction, the phase
shift increases and reaches the maximum value $\Delta\varphi=2\pi$
when the vortex is at the edge of the junction, $\Theta_v=\pi$.
The doubling of the phase shift with respect to the polar angle is
due to concentration of current in the narrow gap between the
vortex and the junction. It can be viewed as being due to an
additional currents of the same sign from the image antivortex on
the other side of the junction. This doubling is quite peculiar.
The total phase shift around the vortex is always $2\pi$. However,
at $z_v \rightarrow 0$ all of it is concentrated at one point at
the nearest edge. This is, apparently, inconsistent with a naive
picture of a rigid phase rotation around the vortex. Curiously, if
we would make junctions at all the edges of vortex-carrying
electrode, all other junctions except the one at which the vortex
is placed would not show any phase shifts in this case. Below we
will perform a similar experiment - simultaneous detection of the
phase shift from different sides of the vortex. This, however,
could be done only for a finite-size electrode. For a finite $L_z$ reflections from the top edge of the electrode leads to appearance
of the top image anti-row II. Subsequent reflections from bottom
and top edges lead to appearance of an infinite series of image
rows I, I', II, II', III, e.t.c, as sketched in Fig. \ref{fig:T2}
(a) (see Appendix A for details).

\begin{figure*}[t]
    \includegraphics[width=0.99\textwidth]{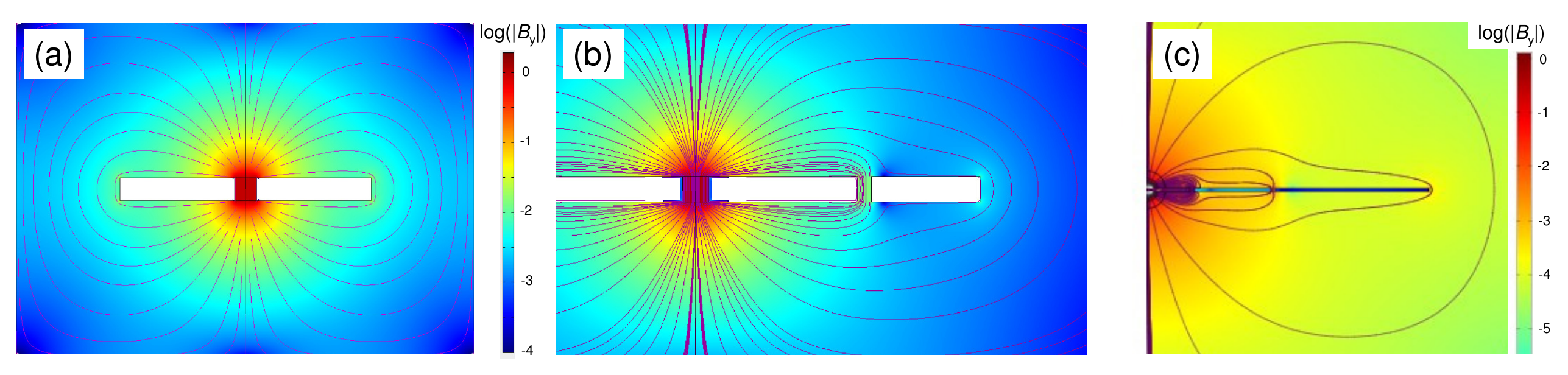}
     \caption{\label{fig:T3} (Color online). 3D numerical simulation of stray field distribution (in a logarithmic scale) from a vortex in (a) one electrode;
     (b) two electrodes, forming a single junction; and (c) three electrodes forming two planar junctions. From panels (b) and (c)
     it can be seen that the stray field lines from the vortex are predominantly closing through the nearby junction and that
     only few are reaching the outmost junction. This is consistent with the cutoff phenomenon observed in experiment. }
\end{figure*}

Inset in Fig. \ref{fig:T2} (c) shows a sketch of a double-junction
experiment, which we will present below. In this case the vortex
is placed in a rectangular electrode with two junction on top and
bottom edges. Solid lines in Fig. \ref{fig:T2} (c) show phase
shifts in the bottom junction versus polar angle of the antivortex
in the bottom junction for square $L_z=L_x$ (blue) and rectangular
$L_z=0.5 L_x$ (red) electrodes. Dashed lines represent
corresponding phase-shifts in the top junction. It can be seen
that upon moving the antivortex from one junction to another the
phase shift changes from $2\pi$ to 0 in the former and from 0 to
$2\pi$ in the latter.

The uneven, not proportional to the polar angle, Eq. (\ref{AQ1}),
variation of the phase shift is characteristic for the considered
mesoscopic case and is associated with non-linearity (singularity)
of current distribution around the vortex center,
Eq.(\ref{J_Pearl}). Therefore, this disproportionality can be
considered as a signature of the circulating current mechanism of
vortex-induced Josephson phase shift. Finally, we note that
although the phase shift from the vortex in this limit is
determined by Eqs.(\ref{AQ1},\ref{J_Pearl}) and decays in a
long-range manner $\propto 1/r$, this is not a long range
phenomenon because it is valid only at $r\ll\lambda^*$. In
principle, the image method can be extended to $r>\lambda^*$
provided there are no stray fields
\cite{Aslamazov_1984,Gurevich_1994,Mironov_2017}. This is the case
for infinite-size objects with zero demagnetization factor. But we
will not discuss it further because it is not relevant for the
considered case of planar thin-film junctions
\cite{Golod_2010,Golod_2015} and because this effect is rapidly
decreasing at $r>\lambda^*$ \cite{Aslamazov_1984} and becomes
insignificant at longer scales.

\subsection{II B. Stray-field effect in a large junction $L_x\gg\lambda^*$}

Next we consider an opposite limit, $L_x\gg\lambda^*$, which is
usually the case for Josephson junctions. Now the vortex carries
the full flux quantum and magnetic field effects are essential. If
vortex-junction distance is (much) larger than the effective
penetration depth $z_v>\lambda^*$, we may neglect circulating vortex currents in the bulk of electrodes 
at the junction. In this case vortex stray
fields become dominant due to a large demagnetization factor of
planar junctions \cite{Golod_2019}. Fig. \ref{fig:T1} (a) and (b)
represent sketches of the top and side views of a planar junction
with an Abrikosov vortex in electrode-1. Closed lines with arrows
in (b) represent stray-field lines and in (a) streamlines of stray
field-induced {\it surface} currents. Since stray fields can not
enter the superconductor, they have to spread along the surface
until the edge up to arbitrary long distance, thus creating truly
long range phenomena, determined by electrode size, rather than
$\lambda^*$.
Distribution of long-range surface currents can be understood
using the short-circuit principle: ``In a complex system of
superconducting films, if two neighboring film surfaces are
short-circuited, it will not affect the current distribution in
any other part of the system other than the two short-circuited
surfaces" \cite{Schmidt}. It means that at a long range current
distribution in electrodes would be the same as if there were no
junction. Therefore, outmost dashed current streamlines in the top
and bottom electrodes in Fig. \ref{fig:T1} (a) should complete the
same circle.

When vortex stray field reach the junction, the corresponding
fraction of the flux closes through the junction, as sketched in
Fig. \ref{fig:T1} (b). This creates an actual magnetic field in
the junction with the sign opposite to that of the vortex. This
field is large because of flux-focusing effect (large
demagnetization factor) \cite{Golod_2019}. It induces
correspondingly large edge currents $I_{1x,2x}$ in the junction
banks, which have equal amplitudes but opposite directions,
$I_{1x}=-I_{2x}$, as sketched in Figs. \ref{fig:T1} (a,b). We want
to emphasize the principle difference between circulating currents
of the vortex, which are bulk (flow in the whole film thickness)
and short-range $r\sim \lambda^*$, and stray-field induced surface
currents, which are long range determined by the geometry of
electrodes. In Fig. \ref{fig:T1} (b) we tried to separate them by
painting short-range bulk vortex currents in white and long-range
stray-field-induced currents in orange/blue. In the limit
considered here, $L_{x,z} \gg \lambda^*$, only stray-field-induced
surface currents are present at the junction banks.

The phase shift, induced by the stray field is given by Eq.
(\ref{AdphiF}), where $\Phi$ is the total stray flux closing
through the junction. Since electrodes have a large area,
$L_{x,z}\gg\lambda^*$, all the field reaching the junction goes in
it because the penalty for stretching it further is too high. The
stray flux can be calculated exactly for the Corbino disk
electrode geometry, see Fig. \ref{fig:T1} (c), with a junction
being a segment of the circle (painted yellow). In this case, due
to rotational symmetry the stray flux in the junction is
\begin{equation}\label{Phi_stray}
\Phi=\frac{\Theta_v}{2\pi}\Phi_0,
\end{equation}
leading to the phase shift, given by the polar angle
\begin{equation}\label{Delta_Phi2}
    \Delta\varphi_v=\Theta_v,
\end{equation}
consistent with Eq. (\ref{AQ1}).

\begin{figure*}[t]
    \includegraphics[width=0.95\textwidth]{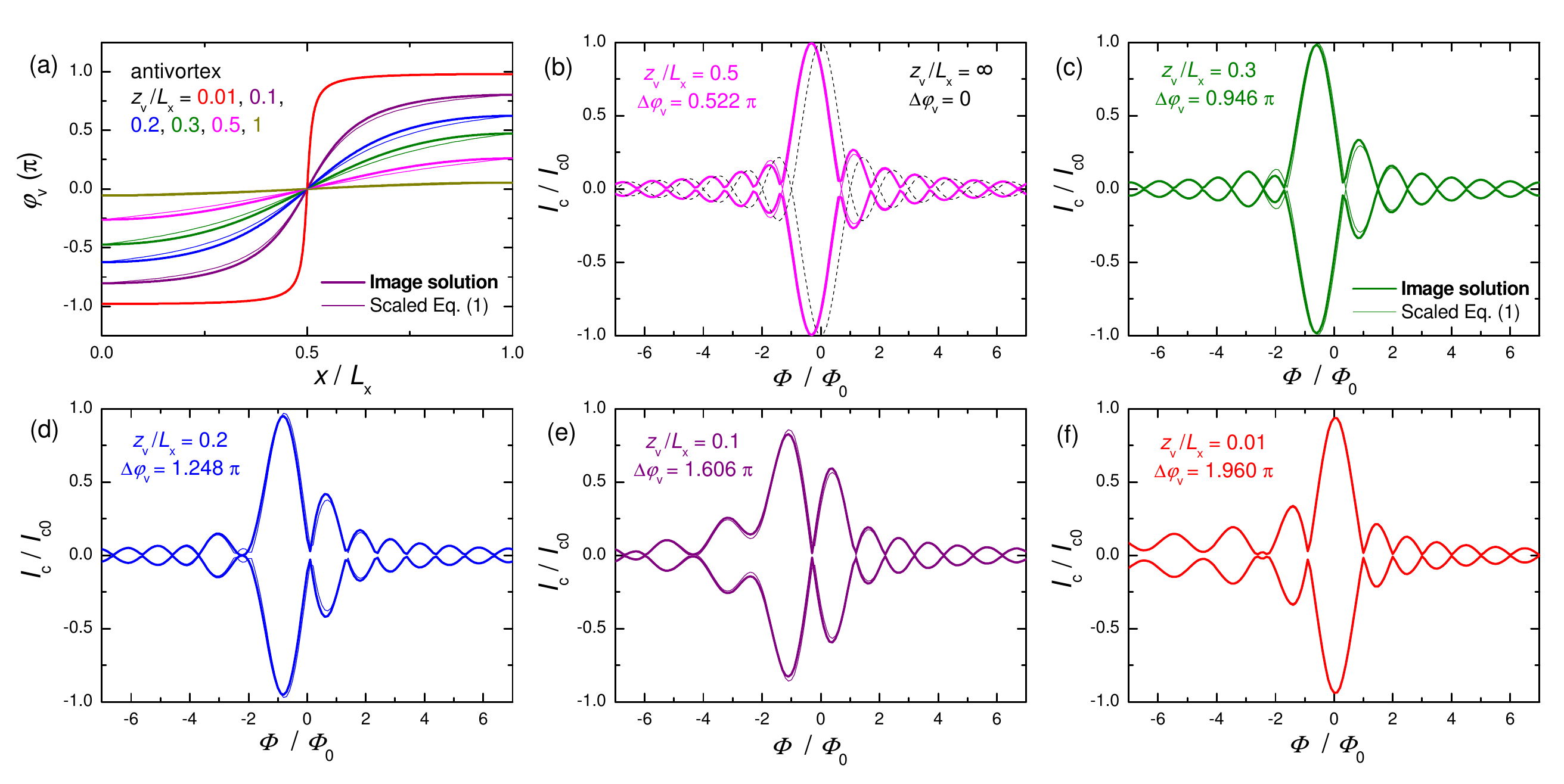}
     \caption{\label{fig:T4} (Color online). Numerical modelling of vortex-induced distortion of $I_c(H)$ patterns. Simulations are
     made for an antivortex at different distances $z_v$ along the middle line $x_v=L_x/2$ of a long $L_z=\infty$ electrode. (a)
     Phase shifts calculated from image array solution (thick lines) and Eq. (\ref{AQ1}), scaled to the same amplitude $\Delta\varphi_v$(thin lines).
     (b-f) Variation of $I_c(H)$ patterns upon approaching of the antivortex to the junction. Dashed black line in (b) represents
     the vortex-free Fraunhofer pattern. Thick and thin lines in (b-f) are obtained for corresponding lines from (a). }
\end{figure*}

Finding distribution of currents and stray fields in realistic
planar junctions with large demagnetization factors is a complex
three-dimensional problem, which can be solved only numerically.
Figure \ref{fig:T3} represents corresponding simulations
\cite{Note1}. We represent data in logarithmic scale so that color
scales represent $\log_{10}(|B_y|)$ and densities of field lines
$\log_{10}(|B|)$. Fig. \ref{fig:T3} (a) shows stray field
distribution from a vortex in the middle of the square shape
superconducting film. In Fig. \ref{fig:T3} (b) additional
electrode is added to the right of the initial electrode with the
vortex. The gap between the two electrodes represents the
junction. It is seen that upon reaching the junction the stray
field is mostly penetrating into it and only a very small fraction
reaches the right edge of the junction. A simulation in Fig.
\ref{fig:T3} (c) represents the case with a third wider electrode
added to the right, thus forming two junctions. It can be seen
that most of the stray field lines from the vortex are closed in
the nearest junction and very few are expanding further out (note
the logarithmic scale). The incomplete closing of stray fields in
the nearest junction in presented simulations is a consequence of
a too small demagnetization factor (scaling with $L_{x,z}/d$) that
we had to adopt for making a reasonably coarse mesh, to solve the
problem on a personal computer. Planar junctions, studied below
have a much larger $L_{x,z}/d \sim 100$, with proportionally
larger energetic penalty for stretching stray field lines to the
outmost edge. Thus, from the 3D simulations, we anticipate that
although the stray field could stretch to an arbitrary long
distance along the electrode, it would be terminated by the first
interruption junction.

We have checked that the stray flux and the phase shift in the
junction indeed scales approximately proportionally to the polar
angle $\Theta_v$, qualitatively consistent with Eq. (\ref{AQ1}).
This indicates a uniform radial spreading of vortex stray-fields.
Therefore, in the considered limit of a large junction $L_{x,z}\gg
\lambda^*$ and for $z_v \gg \lambda^*$ the phase shift is induced
solely by stray fields, proportionally to the polar-angle Eq.
(\ref{AQ1}).

To summarize this section, there are two distinct contributions to
vortex-induced Josephson phase shift: (i) A short-range
$r\lesssim\lambda^*$ circulating-current-induced mechanism with a
phase shift determined by the image series solution,
disproportional with respect to $\Theta_v$; (ii) A long range
$r\gg\lambda^*$ stray field mechanism, proportional to $\Theta_v$.
Despite quantitative differences, phase shifts for both mechanisms
appear to be qualitatively similar and decay in a long-range
manner as $1/r$. This is demonstrated in Figure \ref{fig:T4} (a),
which shows vortex-induced phase shifts for the two mechanisms for
an antivortex at different distances $z_v$ from a junction along
the middle line $x_v=L_x/2$ of an infinitely long electrode
$L_z=\infty$. Thick lines represent an image solution (Eq.
(\ref{DeltaPhi_II'}) from Appendix A, which coincides with the
Pearl-vortex solution Eq. (\ref{Clem}) from Appendix B). Thin
lines represent the stray field contribution equal to the polar
angle, Eq. (\ref{AQ1}), scaled to the same total phase shift.
Apparently, the curves have similar shapes with only a minor
difference at the edges due to different boundary conditions. The
circulating current mechanism requires zero phase gradient at the
edges because there is neither current through the edge, not
magnetic flux in the corresponding mesoscopic limit. For the
stray-field mechanism there is a finite field in the junction,
which leads to a finite gradient at the edges, according to Eq.
(\ref{Adphi0}). Both effects are terminated at the junction
because neither circulating currents nor stray fields would
propagate beyond it. Therefore, the main difference and the only
way to discriminate the two mechanisms are different values and
functional dependence $\Delta \varphi (\Theta_v)$ as seen from
comparison of dotted black (stray field) and solid green
(circulating current) lines in Fig. \ref{fig:T2} (b). In the stray
field case the maximal phase shift is equal to the maximal
possible polar angle $\pi$. However, in the circulating-current
mechanism $\Delta \varphi (\Theta_v)$ is nonlinear and can be
doubled to $2\pi$.

\begin{figure*}[t]
    \includegraphics[width=0.7\textwidth]{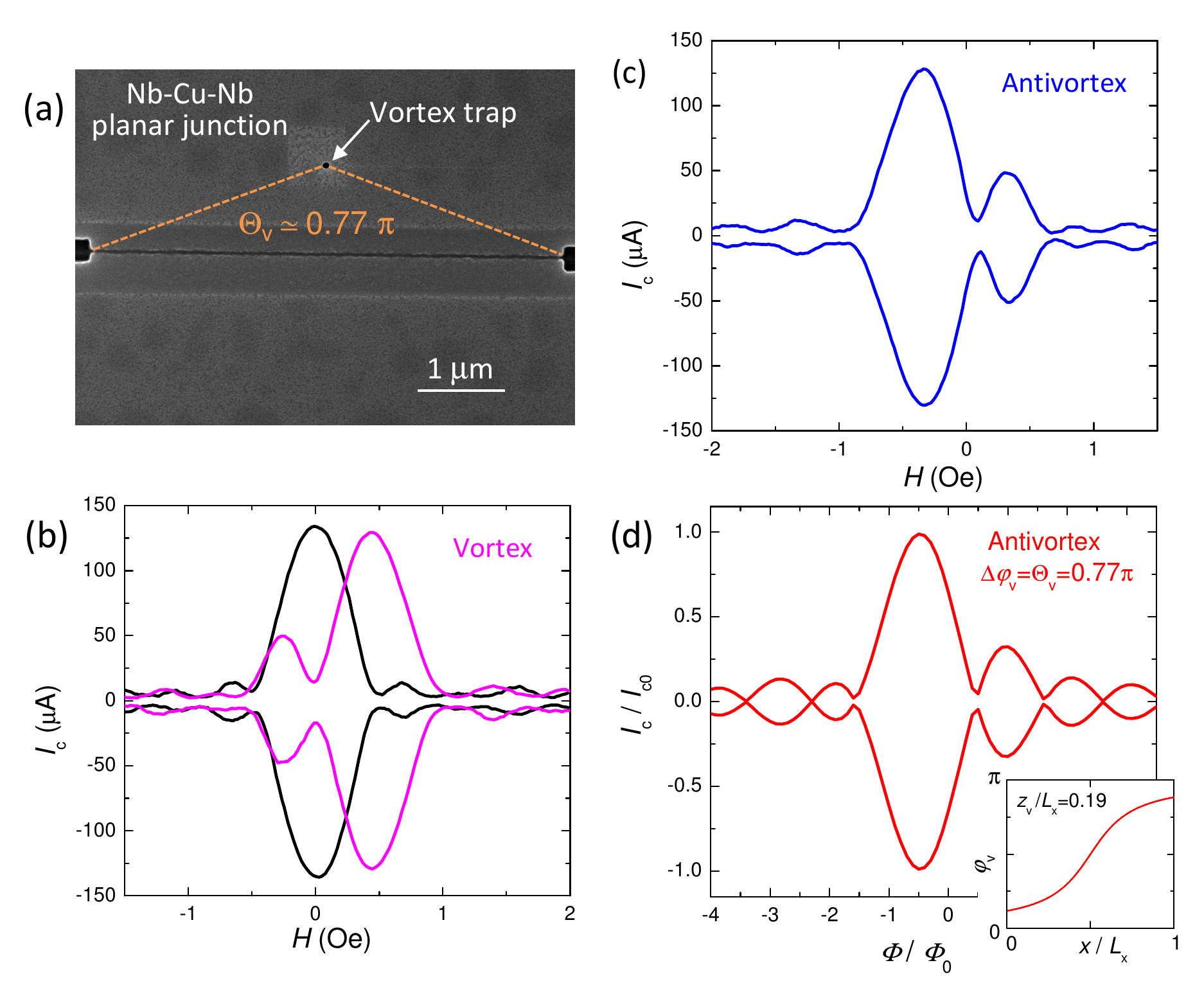}
     \caption{\label{fig5:E1} (Color online). Single junction experiment.
     (a) SEM image of a Nb-Cu-Nb planar junction with a vortex trap in the top electrode at $z_v=1.0~ \mu$m from the
     junction, $z_v/L_x=0.19$ and the polar angle $\Theta_v=0.77\pi$.
     (b) Measured $I_c(H)$ modulation without vortices (black lines) and with a trapped vortex (magenta lines).
     (c) $I_c(H)$ modulation with a trapped antivortex. It is mirror-symmetric with respect to the vortex case in (b).
     (d) Calculated $I_c(H)$ pattern for an antivortex-induced phase shift given by Eq.(\ref{AQ1}) for the experimental
     geometry, shown in the inset. A good agreement with the corresponding pattern in (c) without
     any fitting indicates that the vortex-induced Josephson phase shift is equal to the vortex polar angle, $\Delta\varphi_v=\Theta_v$. }
\end{figure*}

\subsection{II C. Numerical analysis of $I_c(H)$ patterns}

Appearance of vortex-induced phase shift in the junction leads to
distortion of the $I_c(H)$ modulation pattern
\cite{Golod_2010,Clem_2011,Golod_2015}. In a short junction limit
the distortion can be easily calculated by
finding a maximum of Josephson current $I=I_c\sin(\varphi)$,
integrated over the junction length, with $\varphi$ given by Eq.
(\ref{phi_tot}). Successive distortion of $I_c(H)$ patterns upon
approaching the vortex to the junction for phase shifts from Fig.
\ref{fig:T4} (a) are shown in Figs. \ref{fig:T4} (b-f). Thick and thin lines represent image solution and polar angle Eq.(\ref{AQ1}), scaled to the same total phase shift. 
It is seen that the difference between the two solutions is insignificant. Therefore, in the
following we will use Eq. (\ref{AQ1}), with $V$ as a fitting
parameter, for determination of vortex-induced phase shifts from
experimental $I_c(H)$ patterns.

\section{III. Experimental analysis}

\begin{figure*}[t]
    \includegraphics[width=0.99\textwidth]{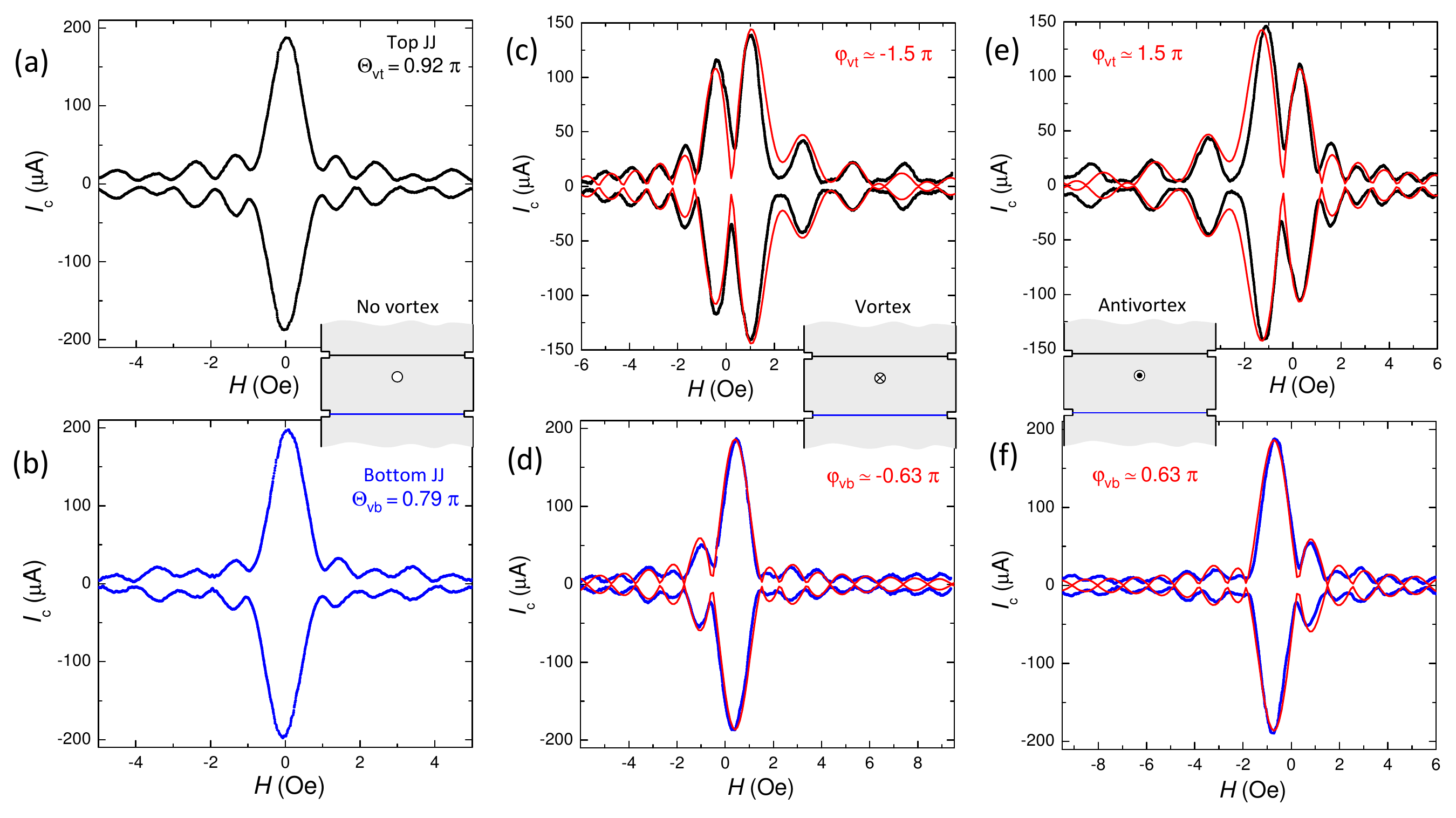}
     \caption{\label{fig6:E2} (Color online). Experiment on a double junction Nb-CuNi-Nb sample with a vortex trap in the middle electrode.
     (a) and (b) Fraunhofer-type vortex free $I_c(H)$
     patterns for (a) top and (b) bottom junction. (c-f) Black and blue symbols represent measured $I_c(H)$
     patterns after trapping (c,d) a vortex (e,f) an antivortex. Note that distortion is larger for the top junction (c,e) which is closer to the vortex.
     Red lines represent fits using Eq. (\ref{AQ1}). Insets show sketches of experiments.}
\end{figure*}

\begin{figure*}[t]
    \includegraphics[width=0.95\textwidth]{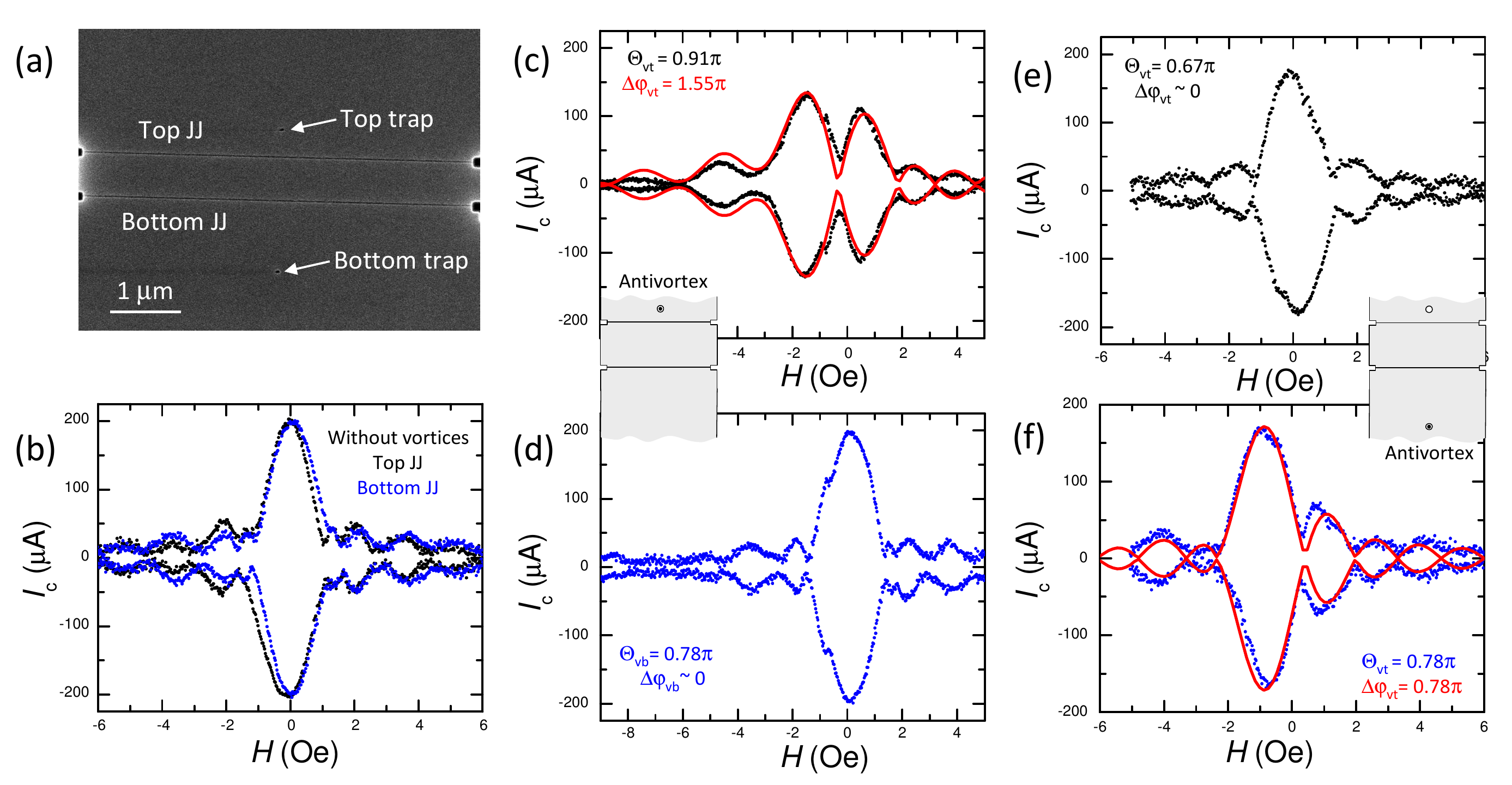}
     \caption{\label{fig7:E3} (Color online). Experimental demonstration of the cutoff effect for a double-junction sample with vortices in outmost electrodes.
     (a) SEM image of the sample with two nearby Nb-CuNi-Nb junctions. The second bottom traps was made after measurements presented in
     (c,d). (b) Vortex-free $I_c(H)$ patterns for both junctions. (c) and
     (d) $I_c(H)$ patterns for the top (c) and bottom (d)
     junctions with a trapped antivortex in the top trap (this is the only trap at this state of the sample, as sketched in the
     inset). Note that the $I_c(H)$ pattern for the top junction is strongly distorted, indicating the large total phase shift
     $\Delta\varphi_v=1.55\pi$ as seen from the numerical fit (red line) in (c). However, the pattern in the bottom junction is practically
     unaffected. (e) and (f) Measurements after making the bottom trap at the same distance to the bottom junction as for the top trap. Symbols represent
     $I_c(H)$ patterns with an antivortex in the bottom trap. This
     time the pattern in the bottom junction is significantly
     distorted, but in the top junction is not. This illustrates
     that the vortex-induced phase shift is terminated at the junction and does not persist beyond it.}
\end{figure*}

Planar proximity-coupled Josephson junctions (of SNS-type, S-
superconductor, N- normal metal) are made by cutting N/S bi-layers
by a Focused Ion Beam (FIB). The bi-layer is deposited by
magnetron sputtering. Films are first patterned into $5-6~ \mu$m
wide bridges by photolithography and ion etching and subsequently
nano-patterned by FIB. Finally, a vortex trap (a hole with a
diameter $30-50$ nm) is made by FIB. We used various N-metals: Cu,
Fig. \ref{fig5:E1}, a superparamagnetic CuNi alloy with
approximately 50-50 concentration, Figs. \ref{fig6:E2},
\ref{fig7:E3}, and some other, all showing similar behavior.
In all presented cases the thicknesses of N-layer (Cu or CuNi) is
50 nm and S-layer (Nb) is 70 nm.  Measurements are done in a
closed-cycle $^4$He cryostat. Magnetic field is applied
perpendicular to the film. More details about fabrication and
characterization of such planar junctions can be found in Refs.
\cite{PhysicaC_2005,Anticorrelation_2007,Golod_2010,Boris_2013,Golod_2015,Golod_2019}.

To study the mechanism of vortex-induced Josephson phase shift we
fabricated devices with vortex traps at different distances $z_v$
from junctions and, correspondingly with different polar angles
$\Theta_v$. The vortex can be manipulated (introduced or erased)
by magnetic field and bias current \cite{Golod_2010,Golod_2015}.
We always start with the vortex-free state, obtained after
zero-field cooling of the sample without bias current. A vortex is
introduced in the trap by applying an appropriate current either at zero field or at small field below the lower critical
field of the electrode, as described in Ref. \cite{Golod_2015}.
Entrance/exit of AV results in an abrupt (instantaneous) change of
the critical current. The two states with and without a vortex in
the trap form steady and perfectly reproducible binary states
\cite{Golod_2015}. This is how we can be sure that the vortex is
indeed sitting in the trap. Only such reproducible states are
shown and analyzed below. It could happen that eventually
additional vortices enter junction electrodes and are placed
randomly outside the trap. However this leads to irreversible and
irreproducible states. If this happens, we clean the device by
repeating the zero-field cooling procedure. For double-junction
experiments, the position of the vortex can be unambiguously
triangulated by simultaneous detection of responses of both
junctions. 

\subsection{III A. Single junction experiment}

Fig. \ref{fig5:E1} (a) shows scanning electron microscope (SEM)
image of a Nb-Cu-Nb planar junction with a vortex trap. The
junction, shown in Fig. \ref{fig5:E1} (a) is made of Cu(50
nm)Nb(70 nm) bilayer. It has a length $L_x \simeq 5.25~\mu$m. The
vortex trap is placed in the middle of the top electrode at a
distance $z_v\simeq 1.0~\mu$m, corresponding to a polar angle
$\Theta_v\simeq 0.77 \pi$. Black lines in Fig. \ref{fig5:E1} (b)
represent measured $I_c(H)$ modulation for this junction at
$T=2.5$ K in the vortex-free state, obtained after zero-field
cooling. It has a regular Fraunhofer-type modulation with a single
central maximum at zero flux, $\Phi=0$.

Magenta lines in Fig. \ref{fig5:E1} (b) represent the measured
$I_c(H)$ pattern for the same junction after trapping a vortex in
a positive field. Apparently, the $I_c(H)$ is significantly
distorted. The main maximum is shifted towards positive field and
an additional secondary maximum appears at the left side. The
shift of the main maximum in the direction of applied field is
very characteristic for a trapped vortex \cite{Golod_2010}. Since
the main maximum corresponds to $\Phi\simeq 0$, such a shift
indicates that the effective vortex-induced field in the junction
is reverse with respect to the applied field. I.e., in Fig.
\ref{fig5:E1} (b) the vortex-induced flux is negative and a
positive field is needed to compensate it to $\Phi=0$, leading to
a corresponding shift of the main maximum $I_c(H)$. This is
opposite to the $I_c(H)$ pattern shift, which occurs in
ferromagnetic junctions \cite{Iovan_2014}. This sign of the
vortex-induced flux is expected for the stray-field, as sketched
in Fig. \ref{fig:T1} (b). Fig. \ref{fig5:E1} (c) shows measured
$I_c(H)$ pattern for the same junction with a trapped antivortex
in a negative applied field. Apparently, it is mirror-reflected
with respect to the vortex case, Fig. \ref{fig5:E1} (b).

In the case depicted in Fig. \ref{fig5:E1}, the vortex is placed
at a rather large distance $z_v=1~\mu$m from the junction, which
is significantly longer than both London, $\lambda \sim 100-150$
nm, and Pearl, $\lambda_P\sim 140-320$ nm, lengths.
Therefore, as explained in sec. II B, vortex stray fields should
make a dominant contribution to the Josephson phase shift and it
should be $\simeq\Theta_v$. Fig. \ref{fig5:E1} (d) represents
numerically calculated $I_c(H)$ with the antivortex-induced phase
shift according to Eq. (\ref{AQ1}) with experimental $\Theta_v
\simeq 0.77 \pi$.
Apparently, the simulation in Fig. \ref{fig5:E1} (d) is in a
quantitative agreement with the experimental data in Fig.
\ref{fig5:E1} (c) without any fitting. From such an analysis we
conclude that the vortex-induced Josephson phase shift scales with
the polar angle, $|\Delta\varphi_v| \simeq \Theta_v$, for large
$z_v \gg \lambda^*$, consistent with earlier report
\cite{Golod_2010}.

\subsection{III B. Double junction experiment}

As discussed in sec. II above, double junction experiment can
provide a crucial information about the mechanism of
vortex-induced phase shift.
We made samples, containing two planar junctions and a vortex trap
at different places. Double-junction devices have a cross-like
geometry with four electrodes: top and bottom electrodes and left
and right electrodes connected to the middle electrode between the
junctions. Thus we can send the current through both junctions
from bottom to top electrode and measure simultaneously
characteristics of both junctions
(for more details see Refs. \cite{Golod_2015,Golod_2019}). Figures
\ref{fig6:E2} (a) and (b) show vortex-free $I_c(H)$ modulations
measured simultaneously for two Nb-CuNi-Nb junctions at the same
device at $T=6.6$ K. Both junctions exhibit regular
Fraunhofer-like modulation $I_c(H)$, indicating good uniformity of
the junctions \cite{Krasnov_1997} and reproducibility of the
fabrication technique.

First we consider the case when a vortex trap is placed in the
middle electrode, shared by both junctions, as sketched in the
inset in Fig. \ref{fig:T2} (c). For this device junction lengths
are similar $L_{xt}=5.4~\mu$m and $L_{xb}=5.43~\mu$m and
separation between junctions is $\simeq 1.3~\mu$m. The vortex trap
is placed closer to the top junction, $|z_{vt}| \simeq 0.36~\mu$m,
and $z_{vb} \simeq 0.94~\mu$m to the bottom junction, as sketched
in the inset. Corresponding polar angles are $\Theta_{vt}\simeq
0.92 \pi$ and $\Theta_{vb}\simeq 0.79 \pi$. In this case the
vortex should simultaneously induce dissimilar phase shifts in
both junctions, as shown in Fig. \ref{fig:T2} (c).

Black and blue symbols in Figs. \ref{fig6:E2} (c) and (d) show
measured $I_c(H)$ patterns for top and bottom junctions,
respectively, after trapping a vortex.
Figs. \ref{fig6:E2} (e) and (f)
show corresponding measurements with a trapped antivortex in
negative field. Apparently, vortex, Figs. \ref{fig6:E2} (c,d), and
antivortex, Figs. \ref{fig6:E2} (e,f), characteristics are mirror
symmetric. From Fig. \ref{fig6:E2} it is obvious that a vortex in
the middle electrode distorts $I_c(H)$ patterns in both junctions,
but the distortion is stronger in the top junction, which is
closer to the vortex, consistent with simulations from Fig.
\ref{fig:T4}. Thus, a double-junction geometry allows an
unambiguous triangulation of the vortex position.

In order to obtain the value of $\Delta \varphi_v$ we perform
fitting of $I_c(H)$ data, as described above. Red lines
in Figs. \ref{fig6:E2} (c-f) represent such fits. From those we
obtain $\Delta\varphi_{vt}\simeq -1.5\pi$, which has a
significantly larger absolute value than $\Theta_{vt}\simeq 0.92
\pi$ and $\Delta\varphi_{vb}\simeq - 0.63 \pi$, with a smaller
absolute value compared to $\Theta_{vb}\simeq 0.79\pi$. This is
inline with calculations for a double-junction
sample with mesoscopic rectangular electrode from Fig.
\ref{fig:T3} (c): when vortex is close to the top junction (small
$\Theta_{vb}$) the phase shift is concentrated in the top junction
and becomes larger than $\Theta_{vt}$ up to a factor two for
$z_{vt}\rightarrow 0$. However, this is accompanied by the
reduction in the furthermost bottom junction, compare dashed and
solid lines in Fig. \ref{fig:T3} (c), qualitatively consistent with our observation.

\subsection{III C. The cutoff effect}

Next we'll consider a specific case when the trap is placed in the
outer electrodes. Figure \ref{fig7:E3} (a) shows a SEM picture of
a double-junction device. Vortex-free characteristics for both
junctions at $T=6.7$ K are shown in Fig. \ref{fig7:E3} (b).
Junction lengths are $L_{xt}=5.58~\mu$m and $L_{xb}=5.56~\mu$m. In
this sample we made junctions close to each other at a distance
only $0.6~\mu$m so that polar angles to both junctions are not
very different. Initially, a single vortex trap was made in the top electrode close to the top junction, marked as the top
trap in Fig. \ref{fig7:E3} (a). The corresponding distances to the
top and the bottom junction are $z_{vt}=0.38~\mu$m and
$z_{vb}=1.0~\mu$m, respectively, and polar angles are almost the
same as in the case of Fig \ref{fig6:E2}: $\Theta_{vt}\simeq
0.91\pi$ and $\Theta_{vb}\simeq 0.78\pi$. However, the result of
vortex trapping is quite different.

Figs. \ref{fig7:E3} (c) and (d) show $I_c(H)$ patterns for the top
and bottom junctions, respectively, with a trapped antivortex in
the top trap at $T\simeq 6.7$ K. It is seen that $I_c(H)$ in the
top junction is strongly distorted. However, the bottom junction
is practically unaffected by the vortex. Red lines in panel (c)
represent a numerically simulated fit, which yields
$\Delta\varphi_{vt}=1.55 \pi$ in the top junction, while in the
bottom junction there is no visible distortion with respect to the
vortex-free case $\Delta\varphi_{vb}\simeq 0$. Apparently, the
vortex in the top electrode does not induce a sensible phase shift
in the bottom junction despite a quite large polar angle
$\Theta_{vb}=0.78\pi$.

After this measurement the sample was taken back to FIB and a
second bottom trap was made in the bottom electrode, marked in
Fig. \ref{fig7:E3} (a), at the same distance and polar angle to
the bottom junction as for the top trap, $|z_{vb}|=1.0~\mu$m,
$\Theta_{vb}=0.78\pi$. Corresponding values for the top junction
are $|z_{vt}|=1.6~\mu$m and $\Theta_{vt}=0.67\pi$. Figs.
\ref{fig7:E3} (e) and (f) show $I_c(H)$ modulation for top and
bottom junctions with a trapped antivortex in the bottom trap. It
is seen that $I_c(H)$ in the bottom junction, which shares one
electrode with the vortex, is now significantly distorted by the
vortex at such a distance, while the top junction, which has
vortex-free electrodes, is unaffected. A numerical fit, shown by
red lines in Figs. \ref{fig7:E3} (f), yields $\Delta\varphi_{vb}=
0.78\pi$, equal to the corresponding polar angle $\Theta_{vb}$.

Observations from Fig. \ref{fig7:E3} clearly demonstrate that the
vortex-induced Josephson phase shift is cutoff by the nearest
junction and does not persist beyond it. In other words, the phase
shift is induced only when the vortex is placed in one of the two
junction electrodes and is not observed otherwise, irrespective of
the distance between the vortex and the junction (for
$z_v>\lambda$). The cutoff effect makes it evident that this is
not the Aharonov-Bohm effect that causes the phase shift. On the
other hand, the cutoff is expected both for current and
stray-field mechanisms, discussed in sec. II A and II B.

The current-induced phase shift cuts-off because the vortex
current does not cross the junction, up to an accuracy given by
the smallness of the Josephson current density $J_c$ compared to
the circulating current density in the vortex $J_v(r)$. In the
mesoscopic case, $r<\lambda^*$ $J_c/J_v$ is very small, which
ensures negligible vortex currents on the other side of the
junction. At large distances circulating currents are small due to
an exponential decay in case of an Abrikosov vortex or quadratic
decay in case of a Pearl vortex \cite{Pearl_1964}, which leads to
an effective cutoff at $z_v\gg\lambda^*$.

For the stray-field mechanism, the cutoff is caused by the
energetic penalty of stretching the magnetic field lines beyond
the nearest junction. It leads to almost complete closing of field
lines through the nearest junction, as confirmed by numerical
modelling in Fig. \ref{fig:T3} (c).

\begin{figure}[t]
    \includegraphics[width=0.4\textwidth]{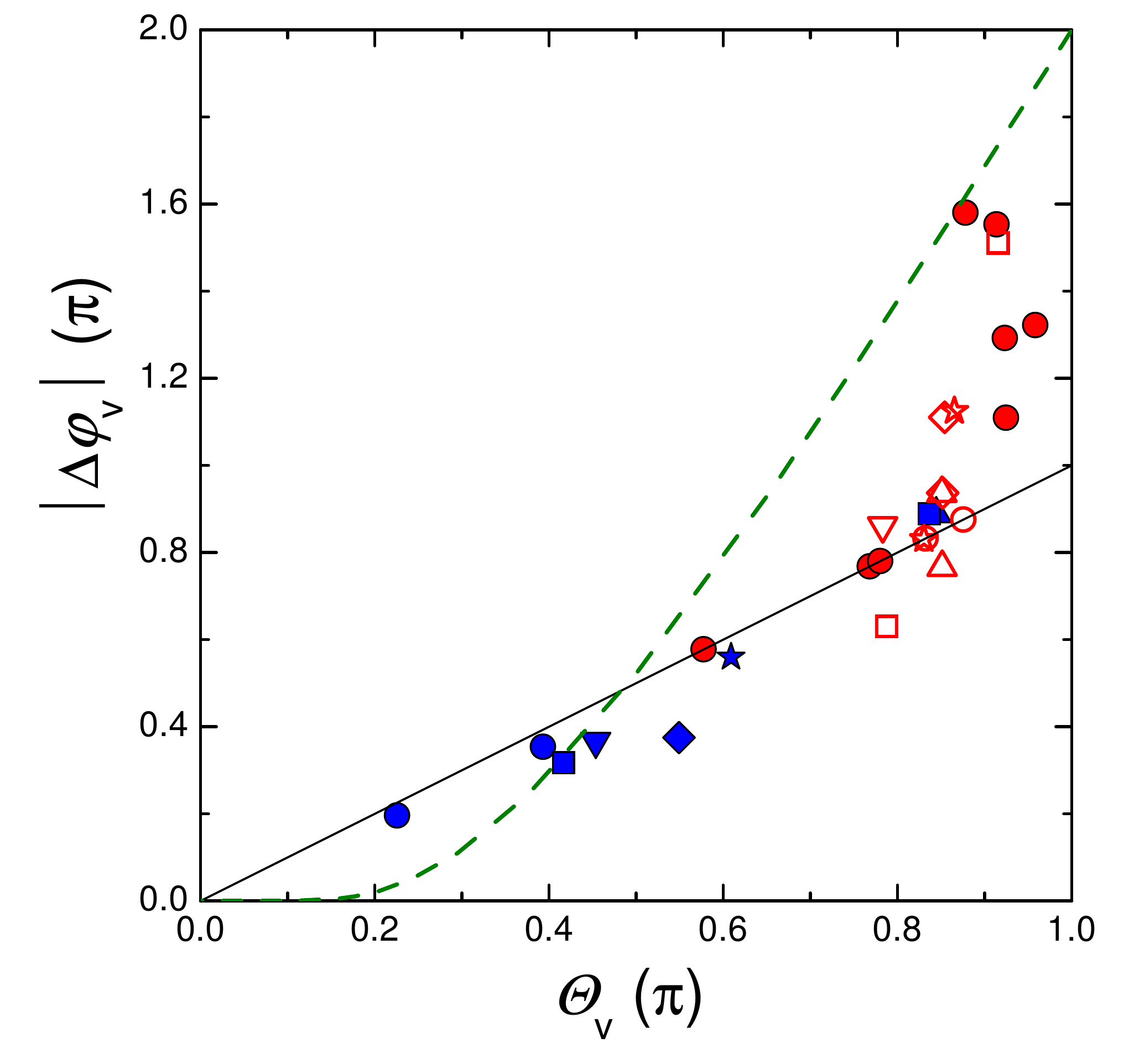}
     \caption{\label{fig8:ES} (Color online). The absolute value of the total vortex-induced Josephson phase shift as a function of
     the vortex polar angle within the junction. Blue symbols represent data from Ref. \cite{Golod_2010}. Red symbols are from present
     study. Solid circles are for the single junction samples and for double-junction samples with vortices in outmost electrodes.
     Open symbols are for double-junction samples with a vortex in the middle electrode. Similar symbols correspond to the two junctions at the
     same sample. Solid and dashed lines represent Eq. (\ref{AQ1}) and image solutions ($L_z=\infty$), respectively.
     A crossover from linear to superlinear dependence occurs at $\Theta_v \gtrsim 0.9\pi$, which corresponds to $z_v\lesssim 500$ nm.
     The crossover indicates the change in the dominant mechanism of the phase shift from stray-field-induced at large distances to
     circulating-current-induced at short distances, comparable to the penetration depth $\lambda^*$. }
\end{figure}

\subsection{III D. Distance dependence of the phase shift}

Figure \ref{fig8:ES} summarizes the dependence of observed phase
shifts on the vortex polar angle. Blue symbols represent
previously reported data from Ref. \cite{Golod_2010}. Red symbols
represent new data from a variety of studied samples. Filled red
symbols correspond to either single-junction samples, see Fig.
\ref{fig5:E1}, or double-junction samples with a vortex in outmost
electrodes, see Fig. \ref{fig7:E3}. Open red symbols represent
double-junction data with a vortex in the middle common
electrodes, as in Fig. \ref{fig6:E2}. Here similar symbols
represent phase shifts in the two junction on the same device. For
example, open squares represent data from Fig. \ref{fig6:E2}. The
values $\Delta\varphi_v$ are obtained from fitting using Eq.
(\ref{AQ1}), as shown in Figs. \ref{fig5:E1}-\ref{fig7:E3}. The
accuracy of determination of $\Delta\varphi_v$ is limited mostly
by some deviations of vortex-free $I_c(H)$ from the pure
Fraunhofer pattern. It may also be affected by presence of
far-away vortices (at $z_v \gtrsim L_x$) that may create a small
irregular distortion of $I_c(H)$, which is difficult to control
just because of its smallness. We estimate the corresponding
systematic uncertainty less than $0.1\pi$.

The straight black line represents the proportional to the polar
angle dependence Eq. (\ref{AQ1}). The dashed green line represents
the image solution 
for a long electrodes $L_z\gg L_x$. From Fig. \ref{fig8:ES} it can
be seen that for small angles (large distances) the vortex-induced
Josephson phase shift scales with the polar angle
$\Delta\varphi_v\simeq \Theta_v$, consistent with earlier report
\cite{Golod_2010}. However at larger polar angles, corresponding
to shorter distances $z_v\lesssim 500$ nm, which is comparable to
the estimated value of the Pearl length $\lambda_P\simeq 300$ nm,
the phase shift becomes larger than the polar angle and approaches
the $2\pi$ value, as predicted by the image solution \cite{Note2}.
We can also see that some of the open red symbols lie below the
$\Delta\varphi_v = \Theta_v$ line. Those points correspond to the
farmost junctions in the double junction case with the vortex in
the middle electrode, as in Figs. \ref{fig6:E2} (d,f). As we
discussed in sec. II A, see Fig. \ref{fig6:E2} (c), such a
deviation is expected for the current-induced mechanism in the
mesoscopic (short-range) case due to concentration of the phase
shift at the opposite junction, nearby the vortex.

The crossover from linear $\Delta\varphi_v\simeq \Theta_v$ to
superlinear $\Delta\varphi_v > \Theta_v$ behavior with decreasing
the distance between the vortex and the junction is consistent
with the theoretical analysis from sec. II and indicates a gradual
transition from a long-range stray-field mechanism at large
distances $z_v\gg\lambda^*$ to a short-range circulating vortex
current mechanism at $z_v\lesssim\lambda^*$. Thus we identify two
distinct mechanisms of Josephson phase shift generation by an
Abrikosov vortex. This is the main result of our work.

\section{V. Conclusions}

To summarize, we performed a systematic study of Abrikosov
vortex-induced phase shift in planar Josephson junctions. We
demonstrated that vortex induced Josephson phase shift
$\Delta\varphi_v$ depends on the polar angle $\Theta_v$ of the
vortex within the junction and decays slowly, in a long-range
manner (non-exponentially), approximately inversely proportionally
to the distance. Thus a significant phase shift can be generated
even at very large distances (few microns), compared to the London
penetration depth $\lambda\simeq 100$ nm. However, experiments
with two consecutive junctions have shown that vortex-induced
phase shift is cutoff by the nearest junction and does not occur
beyond it, irrespective of the distance. Thus, this is not a true
long-range phenomenon, like the Aharonov-Bohm effect. The phase
shift is induced only when the vortex is present in one of the
junction electrodes. On a quantitative level we observed a
crossover from a linear dependence $\Delta\varphi_v\simeq
\Theta_v$ at large distances $z_v\gg\lambda^*$ to a superlinear
dependence at shorter distances $z_v\sim\lambda^*$. To clarify the
origin of vortex-induced phase shift we performed theoretical and
numerical analysis of two mechanisms: circulating vortex currents
at a short range $r\lesssim \lambda^*$ and vortex stray-field at
long range $r\gg\lambda^*$.

For the current-induced mechanism we derived a simple image array
solution in the mesoscopic limit $L_{x}\ll\lambda^*$. In case of
an infinitely long narrow electrode it coincides with the
corresponding Clem's solution for a Pearl vortex \cite{Clem_2011}.
However, the image array solution is valid also for finite-size
electrodes and both for Abrikosov and Pearl vortex. We emphasize,
that the two phase-shift mechanisms are distinct and independent.
The vortex stray field can spread to arbitrary long distances {\it
outside} the superconducting film and can generate the phase shift
at large distances where there are no circulating vortex currents
$\it inside$ the superconductor. Similarly, circulating vortex
currents diverge with approaching the vortex center and create a
large (up to $2\pi$) phase shift only in the short-range
$r<\lambda^*$, where circulating current densities are large. In
the mesoscopic case it occurs without involvement of magnetic
field. Observation of linear-to-superlinear crossover
$\Delta\varphi_v$ versus $\Theta_v$ clearly demonstrates the
gradual transition from one mechanism to another. Altogether this
work provides a comprehensive quantitative description of the
vortex induced Josephson phase shift. We anticipate that it can be
employed for development of future generation of compact Josephson
electronic devices like memory \cite{Golod_2015} and tunable phase
shifters. Both the compactness and the tunability are facilitated
by the small size of the Abrikosov vortex $\lambda\sim 100$ nm,
which represent the smallest magnetic object in a superconductor.

\begin{acknowledgments}
Acknowledgments: The work was accomplished during a visiting
professor semester of VMK at the Moscow Institute of Physics and
Technology and supported by the Russian Science Foundation, Grant
No. 19-19-00594.
\end{acknowledgments}

\appendix

\section{APPENDIX A: Calculation of a phase shift from a vortex-image array in the mesoscopic case $L_x\ll\lambda^*$}

Lets consider a vortex ($V=1$) at position $(x_v, z_v)$ in the
electrode-1 with sizes $(L_x, L_z)$, as sketched in Fig.
\ref{fig:T2} (a). We assume that at least one of the sizes is
small, $L_x\ll\lambda^*$. Mirror reflections from the vertical
(left-right) edges will create a first image row (marked I in Fig.
\ref{fig:T2} (a)) at $z_1=z_v$ with anti-vortices at
$x_{n-}=-2(n-1)L_x - x_v$ and $ x_{n+}=2nL_x - x_v,~ (n=1,2,3...)$
and vortices at $x_{m-}=-2mL_x + x_v$ and $x_{m+}= 2mL_x + x_v,~
(m=1,2,3...)$. Each of them induces a phase shift according to Eq.
(\ref{AQ1}). The original vortex will create a total phase shift
$\Delta\varphi_0=\arctan[(L_x-x_v)/z_v]+\arctan[x_v/z_v]$. The two
primary anti-vortex images (marked green in Fig. \ref{fig:T2} (a))
will reduce it by
$\Delta\varphi_1=-\arctan[(L_x+x_v)/z_v]+\arctan[x_v/z_v]-\arctan[(2L_x-x_v)/z_v]+\arctan[(L_x-x_v)/z_v]$.
Subsequent vortex and antivortex pairs will add
$\Delta\varphi_m=\arctan[((2m+1)L_x-x_v)/z_v]-\arctan[(2mL_x-x_v)/z_v]+\arctan[(2mL_x+x_v)/z_v]-\arctan[((2m-1)L_x+x_v)/z_v]$
and
$\Delta\varphi_{m+1}=-\arctan[((2m+1)L_x+x_v)/z_v]+\arctan[(2mL_x+x_v)/z_v]-\arctan[(2m+2)L_x-x_v)/z_v]-\arctan[((2m+1)L_x+x_v)/z_v]$,
$(m=1,2,3...)$, correspondingly. As a result, the total phase
shift induced by the row-I can be written as

\begin{widetext}

\begin{equation}\label{DeltaPhi_I}
\Delta\varphi_{\textrm{I}}=2 \left(
\tan^{-1}\frac{x_v}{z_v}-\sum_{n=1}^{\infty}
\left[\tan^{-1}\frac{2nL_x-x_v}{z_v}-\tan^{-1}\frac{(2n-1)L_x-x_v}{z_v}-\tan^{-1}\frac{2nL_x+x_v}{z_v}
+\tan^{-1}\frac{(2n-1)L_x+x_v}{z_v} \right]\right).
\end{equation}

\end{widetext}

If electrode-1 is semi-infinite $L_z=\infty$, then we need to add
only one additional image row from the bottom (junction) edge,
marked I' in Fig. \ref{fig:T2} (a). It is obvious that this row
will create exactly the same phase shift, as row-I. Consequently,
for a semi-infinite electrode the total phase shift is double the
phase shift of the primary row.
\begin{equation}\label{DeltaPhi_II'}
\Delta\varphi(L_z=\infty)=2\Delta\varphi_{\textrm{I}}.
\end{equation}

For a finite-size electrode additional image rows appear due to
mirror reflections from the top edge, as depicted in Fig.
\ref{fig:T2} (a). Reflection of the primary row-I from the top
edge leads to the secondary anti-row II $z=2L_z-z_v$. Its
reflection from the bottom edge leads to a row II' at
$z=-2L_z+z_v$, and so on. Each image vortex in the row generates a
phase shift according to Eq. (\ref{AQ1}) with corresponding
coordinates $(x_n,z_m)$. Similar to the case of antivortices and
vortices in the primary row I, antirows appear at
$z_{n-}=-2(n-1)L_z - z_v$ and $ z_{n+}=2nL_z - z_v,~
(n=1,2,3...)$, rows at $z_{m-}=-2mL_z + z_v$ and $z_{m+}= 2mL_z +
z_v,~ (m=1,2,3...)$ and couples of rows are symmetric with respect
to the junction (e.g. rows I - I' and II - II' in Fig.
\ref{fig:T2} (a)) and create identical phase shifts. This
simplifies calculations.

The sum in Eq. (\ref{DeltaPhi_I}) is well behaving and converges
at $n=10-20$, depending on $z_v$. The larger is $z_v$ (small
$\Theta_v$) the more terms are needed.
$n=10$ is enough for achieving an absolute accuracy $\lesssim
0.01\pi$. The simulated curves shown in Fig.\ref{fig:T2} (b) were
obtained with 20 terms and absolute accuracy $\lesssim 0.001\pi$.
For a finite $L_z$, $\Theta_v$ can not be small and, therefore,
convergence is much faster. Data shown in Fig. \ref{fig:T2} (a)
for finite $L_z$ was obtained by summing four pairs of rows, but
already three rows provide similar result with no visible
difference at the scale of the graph.

\section{APPENDIX B: Comparison with the Pearl vortex solution in the mesoscopic case $L_x\ll\lambda^*$}

For a thin film planar junctions with the thickness $d\ll\lambda$
the effective screening length is given by the Pearl length
$\lambda_P=\lambda^2/d\gg\lambda$ \cite{Pearl_1964}, which expands
the range of validity of the mesoscopic limit $L_x\ll\lambda_P$.
In Ref. \cite{Clem_2011} Clem obtained a self-consistent solution
for a phase shift induced by a Pearl vortex in a thin film planar
Josephson junction with narrow, long electrodes using a conformal
mapping method:
\begin{equation}\label{Clem}
\varphi_{Pearl}=\arg \left[
\frac{w(\zeta)-w^*(\zeta_V)}{w(\zeta_V)-w(\zeta)} \right],
\end{equation}
where $\zeta=x+iz$, $\zeta_V=x_V+iz_V$, $w(\zeta) =
i\sinh(\pi\zeta/L)$ and $w^*$ is a complex conjugate of $w$. The
solution is valid for a junction with $d\ll \lambda$,
$L_x\ll\Lambda_P$ and $L_z=\infty$. The dashed magenta line in
Fig. \ref{fig:T2} (b) shows corresponding Pearl-vortex induced
Josephson phase shift. It coincides with the image solution for
the half-infinite electrode $L_z=\infty$, green line in Fig.
\ref{fig:T2} (b). As explained above the phase shift in this case
is twice that for the primary image row I, see Fig. \ref{fig:T2}
(a) and Eqs. (\ref{DeltaPhi_I},\ref{DeltaPhi_II'}).


\end {document}